\documentclass[aps,pra,reprint,superscriptaddress]{revtex4-2}  

\usepackage{graphicx}  
\usepackage{amssymb}   
\usepackage{amsmath}   
\usepackage{hyperref}
\newcommand{\TEMZZ}{TEM$_{00}$}

\begin{document}
\author{David E. Cohen}
\affiliation{Université Paris-Saclay, CNRS/IN2P3, IJCLab, 91405 Orsay, France}
\author{Annalisa Allocca}
\affiliation{Università di Napoli ``Federico II'', Complesso Universitario di Monte S.Angelo, I-80126 Napoli, Italy}
\author{Gilles Bogaert}
\affiliation{Artemis, Université Côte d’Azur, Observatoire Côte d’Azur, CNRS, F-06304 Nice, France}
\author{Paola Puppo}
\affiliation{INFN, Sezione di Roma, I-00185 Roma, Italy}
\author{Thibaut Jacqmin}\email{Corresponding author: thibaut.jacqmin@sorbonne-universite.fr} 
\affiliation{Laboratoire Kastler Brossel, Sorbonne Université, CNRS, ENS-Université PSL, Collège de France, 75005 Paris, France}
\collaboration{Virgo Collaboration}
\noaffiliation
\title{Towards optomechanical parametric instabilities prediction in ground-based gravitational wave detectors}

\date{\today}

\begin{abstract}
Increasing the laser power is essential to improve the sensitivity of interferometric gravitational wave detectors.  However, optomechanical parametric instabilities can set a limit to that power.  It is of major importance to understand and characterize the many parameters and effects that influence these instabilities. Here, we model with a high degree of precision the optical and mechanical modes that are involved in these parametric instabilities, such that our model can become predictive. As an example, we perform simulations for the Advanced Virgo interferometer (O3 configuration).  In particular we compute mechanical modes losses by combining both on-site measurements and finite element analysis with unprecedented level of detail and accuracy.  We also study the influence on optical modes and parametric gains of mirror finite size effects, and mirror deformations due to thermal absorption.  We show that these effects play an important role if transverse optical modes of order higher than four are involved in the instability process.
\end{abstract}

\maketitle

\section{\label{sec:introduction}Introduction}
In 2015, the LIGO-Virgo collaboration~\cite{virgo_website,acernese2014advanced,ligo_website,aasi2015advanced} detected for the first time gravitational waves preceding a binary black hole coalescence~\cite{abbott2016observation}, thus pioneering gravitational-wave astronomy. Today many other gravitational waves have been detected~\cite{abbott2019gwtc,abbott2020gwtc}. These detections have provided confirmation on the expected rate of binary black hole (BBH) mergers~\cite{abbott2016astrophysical}, a better understanding of BBHs population~\cite{abbott2016astrophysical,abbott2019binary}, a better limit to the mass of the graviton~\cite{abbott2017gw170104}, a first direct evidence of a link between binary neutron star (BNS) mergers and short gamma-ray bursts~\cite{abbott2017gw170817}, a higher precision in constraining the Hubble constant~\cite{ligo2017gravitational}, and a better understanding of BNS mergers~\cite{abbott2017gw170817}. Since the first detections, improvements performed on ground based detectors yielded better detector sensitivities. Gravitational wave sources that are weaker or located further away can now be detected. Among the many improvements, increasing the light intensity in the interferometer arm cavities reduces the impact of the laser quantum phase noise, which is limiting the sensitivity in the high-frequency range. However, a laser power increase can trigger a nonlinear optomechanical effect~\cite{braginsky2001parametric,vyatchanin2012parametric}, known as optomechanical parametric instability (OPI). This effect can jeopardize the interferometer stable operation.

During the Observing Run~1 (O1), LIGO experienced an OPI for the first time~\cite{evans2015observation}: after a few seconds, the interferometer went out of lock, thus preventing further data acquisition. In this letter, we present the models that we used to compute the OPI gains for a power recycled gravitational wave detector. Compared to previous work ~\cite{evans2010general}, we provide a precise description of optical and mechanical modes, together with a study of the impact of losses and the thermal deformation of the mirrors. This will allow to perform parametric instabilities predictions, which is of major importance for future designs of ground-based detectors. These simulations are done for the power recycled Advanced Virgo interferometer (O3 configuration), and can be extended to any other configuration. 

\begin{figure}
	\centering
	\includegraphics[scale=0.03]{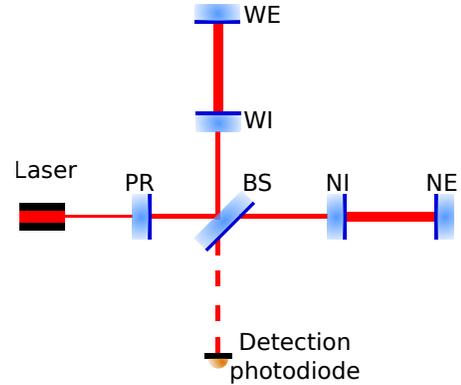}
	\caption{\label{fig:virgo_o3}O3 Advanced Virgo's configuration. NI and NE are respectively the input and end-mirror of the North-arm, and WI and WE are respectively the input and end-mirror of the West-arm. PR is the Power Recycling mirror, BS the interferometer beam splitter. }
\end{figure}

In sec.~\ref{sec:pi}, a short introduction to the model used to compute the parametric gains is given. In sec.~\ref{sec:mms}, we report on a detailed finite element analysis (FEA) of the mirrors, to compute precise mechanical modes frequencies and amplitudes, and estimate quality factors. An original method is then used to combine these FEA simulations with ring-down measurements performed on a subset of modes, in order to obtain accurate quality factors for all the modes.
 In section~\ref{sec:oms}, different models for optical modes are compared: the analytical solution of the paraxial equation for purely spherical infinite size mirrors, as implemented in ~\cite{evans2010general}, and a brute force numerical simulation which includes finite size effects and arbitrary mirror surface shapes. In section~\ref{sec:thermal_effects}, we study the influence on optical modes of a thermal effect related to a local temperature increase of the mirror surface due to light absorption. Finally in section~\ref{sec:pi_computation}, we provide an example of parametric gains that are obtained  in the Advanced Virgo O3 configuration, including optical losses and mechanical losses calculated with an unprecedented level of precision, and we investigate for the first time the effect of the mirror thermal deformation due to laser absorption.

\section{\label{sec:pi}Optomechanical parametric instability}
In an optomechanical cavity like one arm of a gravitational wave detector, photons from the optical zeroth order mode can be coherently scattered to a higher order transverse optical mode if a mechanical mode that sets a mirror surface into motion has its frequency $\omega_m/2\pi$ equal to the frequency difference between the two optical modes (modulo the cavity free spectral range). This phenomenon can remove energy from the mechanical mode by annihilating phonons, and scattering photons from the zero order mode to the higher order transverse mode, thus damping the mirror motion~\cite{Chen}. Conversely, this phenomenon can add energy to the mechanical mode with the reverse process, thus exciting the mechanical motion. In that case, an instability can prevent the interferometer stable operation~\cite{abbott2016observation, braginsky2001parametric}. This instability has a threshold: it starts to grow as soon as the resonant excitation of the mechanical mode by the radiation pressure force overcomes mechanical losses.

In the following, we use the approach developed by Evans {\it et al.}~\cite{evans2010general} to simulate this effect. In this framework, the whole interaction between the three implied modes (two optical modes and a mechanical mode) is seen as a classical feedback system. This modular approach is well suited, since it can be adapted to many different interferometer configurations with the same analytical formulas. The parametric gain of the mechanical mode $m$ is given by
\begin{equation}
    R_m = \frac{8\pi Q_\textrm{m} P}{M\omega^2_m c \lambda} \sum^\infty_{n=0} \Re[G_n] B^2_{m,n}
    \label{OPIGain}
\end{equation}
where $Q_m$ is the quality factor of the mechanical mode $m$ and $\omega_m$ its frequency, $P$ the arm-cavity optical power, $\lambda$ the optical wavelength, $M$ the mirror mass, $c$ the velocity of light, $G_n$ is related to the scattered field optical gain of the $n^\textrm{th}$ optical mode and encapsulates the interferometer configuration. Finally, $B_{m,n}$ is the spatial overlap integral between the three involved modes. A mechanical mode is amplified if $R_m > 0$ and damped if $R_m < 0$. It becomes unstable if $R_m > 1$. 

\begin{figure}[h!]
	\centering
	\includegraphics[width=1.05\linewidth]{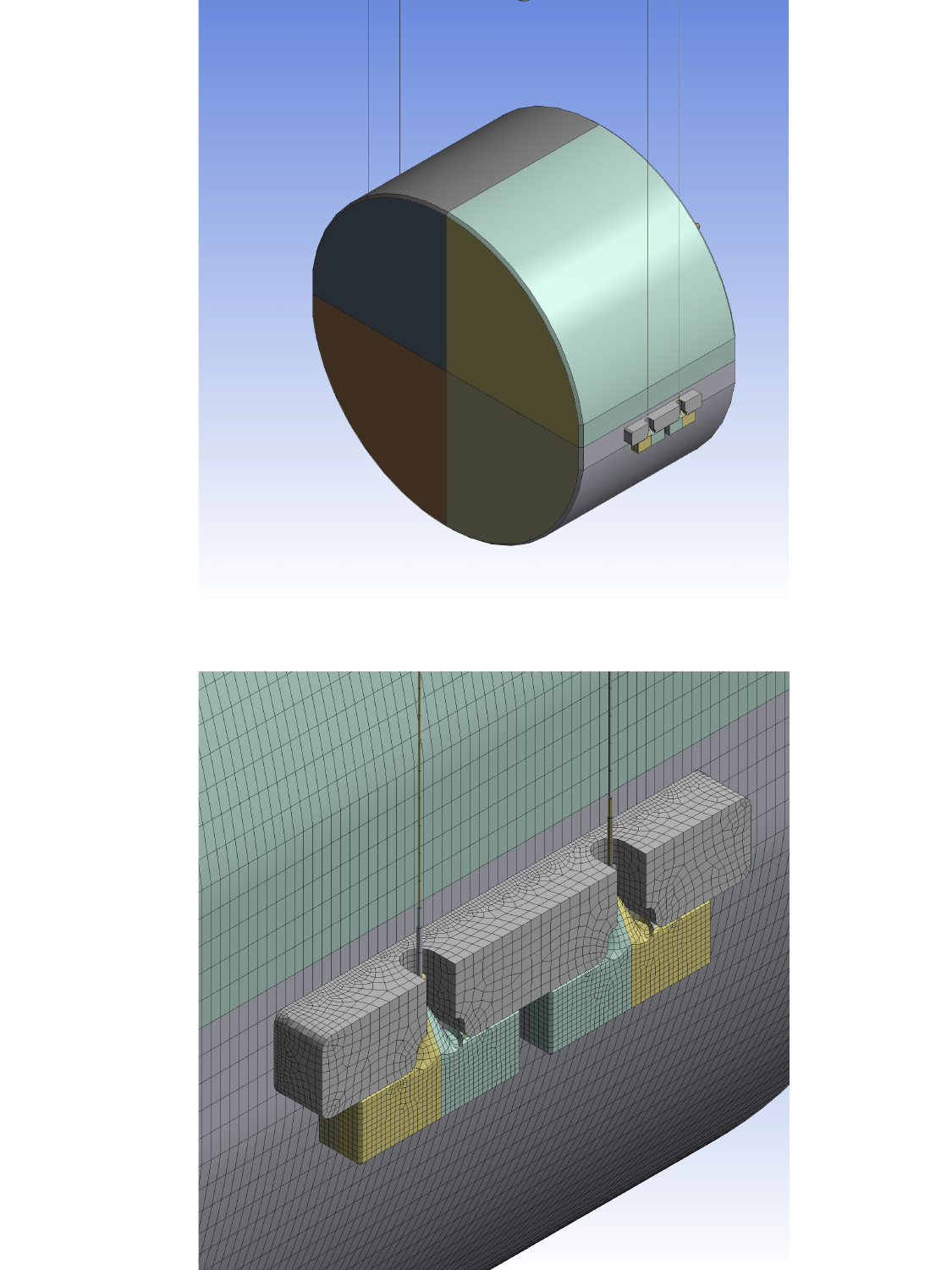}
	\caption{Geometry used for the FEA, including the ears, the anchors and the magnets attached on the mirror rear face. The suspension wires are just for sketching but not included in the simulation as they do not influence the modal frequencies\label{fig:model}}
\end{figure}

\section{\label{sec:mms}Mechanical simulation}

\subsection{The spatial overlap parameter}
The spatial overlap integral $B^2_{n,m}$ is defined~\cite{Grass} as
\begin{equation}
    B^2_{m,n}=\frac{M}{M_\mathrm{eff}}
    \frac{\left(\int E^{00}(\Vec{r})E^{n}(\Vec{r})\mu_\perp^m(\Vec{r})d\Vec{r}_\perp\right)^2}{\int |E^{00}|^2 d\Vec{r}_\perp\int|E^{n}|^2 d\Vec{r}_\perp},
    \label{overlap}
\end{equation}
where $M$ is the mirror mass and $m_{\mathrm{eff}}$ the effective mass of the mechanical mode. The integral is performed over the test mass surface (coating side). $E^{00}(\Vec{r})$ stands for the optical carrier amplitude and $E^{n}(\Vec{r})$ for a transverse optical mode amplitude labeled by the index $n$. As the interferometer is sensitive to the test mass displacement along the optical axis, only the vertical displacement $\mu_\perp^m(\Vec{r})$ is considered, where $m$ is the mechanical mode index. The effective mass is related to the strain energy $\rho_e$ through the equation $\frac{1}{2}M_{\mathrm{eff}}\omega_m^2=\int\rho_e\Vec{r}_\perp$, and effectively obtained with the formula 
\begin{equation}
M_{\mathrm{eff}}=M <\mu(\Vec{r})^2>=M\frac{1}{V}\int{\mu(\Vec{r})^2d\Vec{r}}, 
\end{equation}
where $\mu(\Vec{r})$ stands for the test mass displacement.

 \begin{table}
 \begin{tabular}{|l|c|}
        \hline
        \multicolumn{2}{|c|}{Advanced Virgo FEA parameters}\\
        \hline
        \multicolumn{2}{|c|}{IM Coating\textsuperscript{a}}\\
        \hline
        $Ta_2O_5$ High index &\\ 
        layer overall thickness($t_H^{\text{IM}}$) & 2080~nm\\
        $SiO_2$ Low index &\\ 
        layer overall thickness($t_L^{\text{IM}}$) & 727~nm\\
        Loss angle ($\phi_{\text{CIM}}$)    & $1.1\cdot 10^{-4}f^{0.05}$\\
        \hline
        \multicolumn{2}{|c|}{EM Coating\textsuperscript{a}}\\
        \hline
        $Ta_2O_5$ High index &\\ 
        layer overall thickness ($t_H^{\text{EM}}$) & 3766~nm\\
        $SiO_2$ Low index &\\
        layer overall thickness ($t_L^{\text{EM}}$) & 2109~nm\\
         Loss angle ($\phi_{\text{CEM}}$) & $2.2\cdot 10^{-4}f^{0.01}$\\
        \hline
        \multicolumn{2}{|c|}{TM Suprasil}\\
        \hline
        Young modulus & 72.251~GPa\\
        Poisson ratio & 0.16649 \\
        Density             & 2201~kg.m\textsuperscript{-3}\\  Loss angle ($\phi_{\text{Suprasil}}$)              & $7.6\cdot 10^{-12}f^{0.11}$\\ 
        \hline
        \multicolumn{2}{|c|}{Ear and anchors HCB \cite{cunningham}}\\
        \hline
        Young's modulus               & 72.9~GPa\\
        Poisson ratio               & 0.17  \\
        Density                     & 2201~kg.m\textsuperscript{-3}\\
        Thickness               & 60~nm\\
        Loss angle  ($\phi_{\text{HCB}}$)             & 0.1 \\
        \hline
         \multicolumn{2}{|c|}{IM and EM properties}\\
         \hline
       Mass               & 42~kg\\
      Thickness               & 200~mm\\
       Diameter              & 350~mm\\
       Flats              & 50~mm\\
        \hline
        \end{tabular}
 \caption{\label{FEApar}Mechanical parameters used in the FEA\footnote{\label{notea}For the FEA the multi-layer coating of the IM was replaced by one layer having the total thickness corresponding to the sum of the thicknesses of the high reflective and low reflective materials. The mechanical parameters used are the average values of this layer \cite{granata}}}
 
        \label{Table:mech_par}
\end{table}

The mechanical modes were computed by means of finite element analysis (FEA) developed for the actual input test mass (IM) of Advanced Virgo arm cavities \cite{puppo}. We have used the program  Ansys\textregistered Workbench\textsuperscript{TM}. The IM model includes the high-reflectivity (HR) coating of the front face, the flats and the bevels. Moreover the ears and the anchors attached by silicate bonding technique are included (see figure \ref{fig:model}). In the FEA, the multi-layer optical coating is modelled as a solid 3D element having the total thickness corresponding to the sum of the thicknesses of the high reflective and low reflective materials and mechanical parameters averaged over the thicknesses of the layers. Instead of 3D  {\it shell elements}, we have used 3D {\it solid elements} also for very thin materials, though more CPU time consuming, because they provide the shear deformations and energies, which are useful for getting the mechanical losses associated to the modes. 

\subsection{FEA simulations results}

The flats, the ears and also the anchors play an important role. In particular, since they break the cylindrical symmetry, they lift degeneracies and increase the number of distinct mode frequencies. In this paper we will discuss the results up to 70 kHz.

\begin{figure}[h!]
	\centering
	\includegraphics[width=1.03\linewidth]{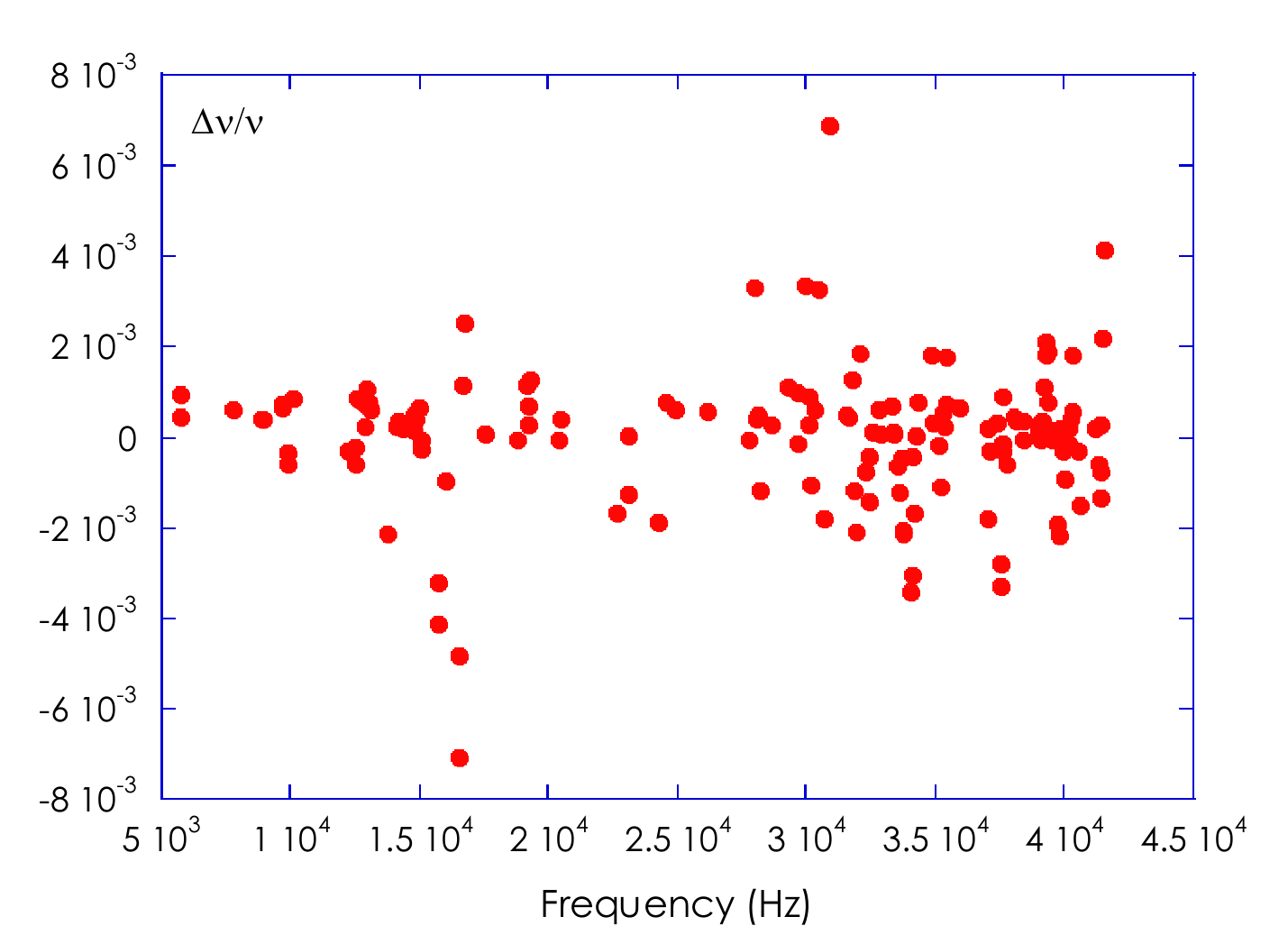}
	\caption{\label{fig:residuals} Relative differences of measured frequencies with respect to frequencies obtained with the FEA vs the FEA frequencies. The standard deviation is $0.15\cdot\%$}
\end{figure}

To estimate the accuracy of the model, we have used a set of frequencies ($\nu_{\mathrm{Meas}}$) measured on the North arm IM up to 40kHz of an IM. 
Fig.~\ref{fig:residuals} shows relative differences $(\nu_{\mathrm{Meas}}-\nu_{\mathrm{FEA}})/\nu_{\mathrm{FEA}}$, versus the frequency of the FEA $\nu_{\mathrm{FEA}}$. The standard deviation is 0.15\%.
v
\begin{figure}[h!]
	\centering
	\includegraphics[width=1\linewidth]{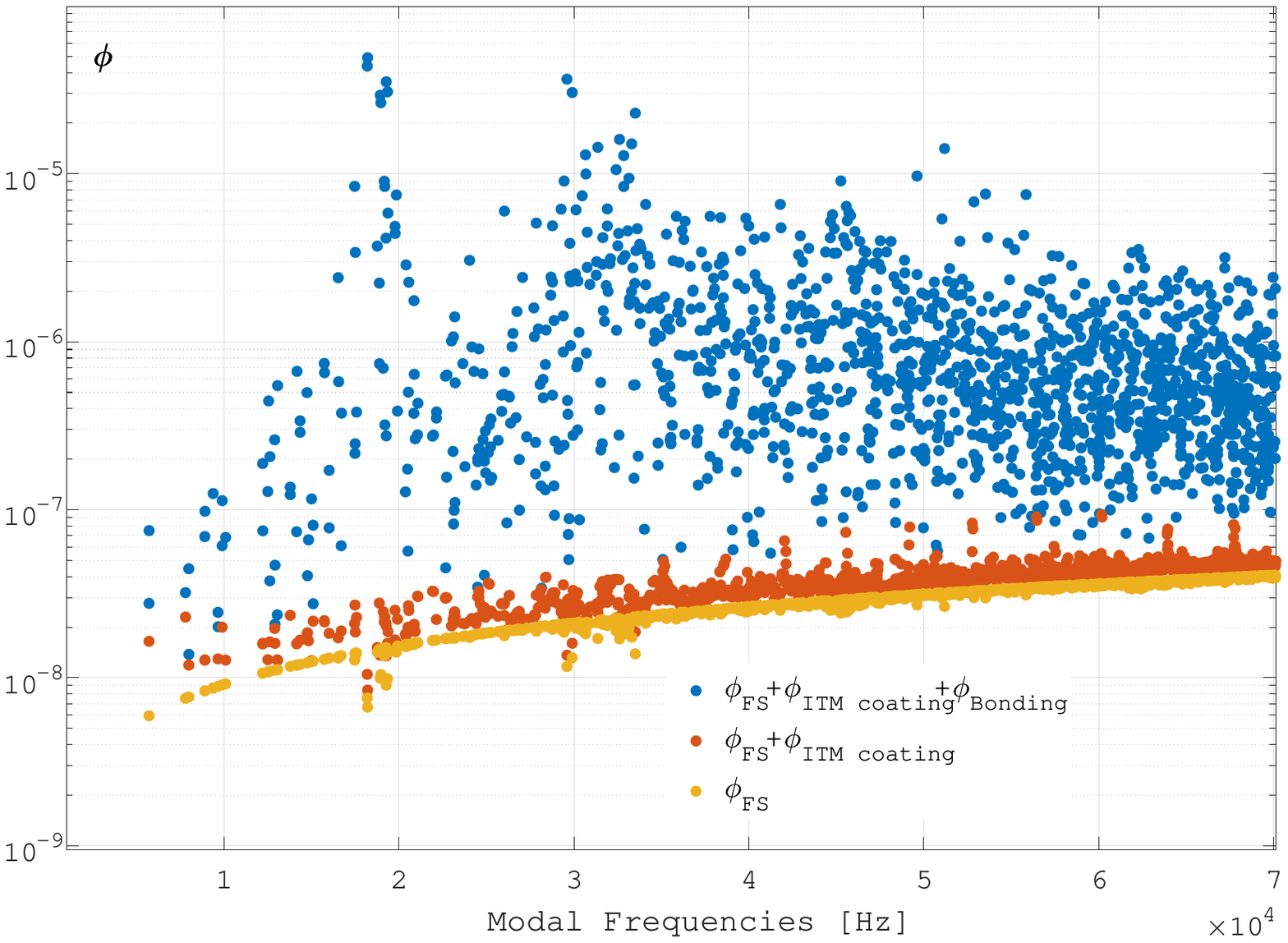}
	\caption{\label{fig:phi_factors} Loss angles obtained from the FEA of the Input TM. The computation ha been performed up to 70 kHz.}
\end{figure}
\begin{figure}[h!]
	\centering
	\includegraphics[width=1\linewidth,angle=0]{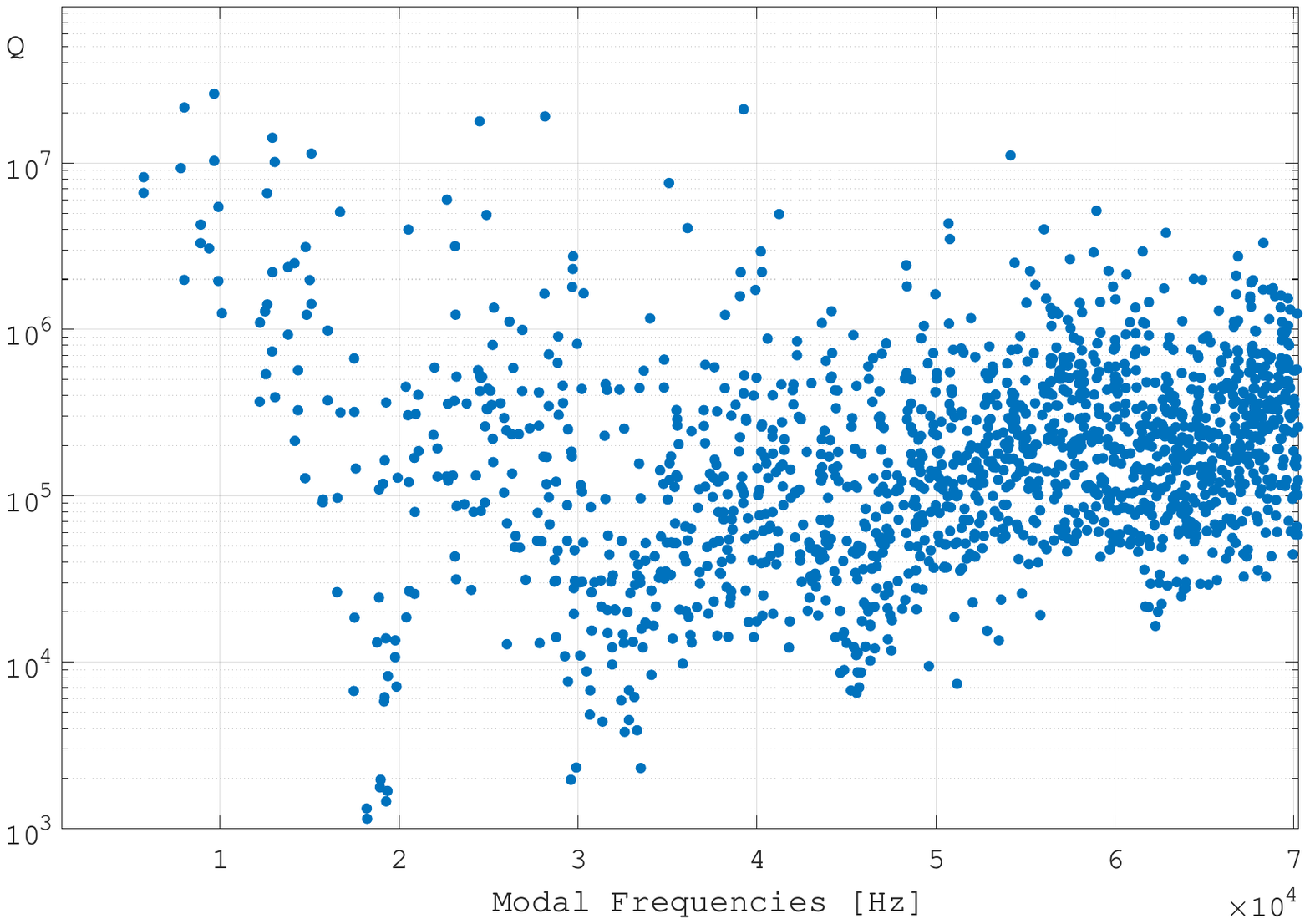}
	\caption{\label{fig:q_factors} Quality factors of the mechanical modes up to 70 kHz}
\end{figure}

We have estimated the quality factors of the mechanical modes of the IM taking into account several kinds of losses: losses of the fused silica substrate, anchors and supports of the magnets (loss angle $\phi_{\mathrm{FS}}$); coating losses (loss angle $\phi_{\mathrm{IM coating}}$); losses of bonding layers used to attach the ears, the anchors and the magnets (loss angle $\phi_{\mathrm{Bonding}}$). The bonding layers, have a thickness of 60~nm, and are modeled as 3D solid elements. Coating losses of the IM and EM were recently measured~\cite{granata}. Note that all the parameters used are given in table \ref{FEApar}. Each loss contributor is related to the energy fraction stored in the lossy part and to the material loss angle, through the relationships
\begin{align}
\begin{split}
   \phi_\mathrm{Bonding}\cdot E_\mathrm{tot}&=\phi_\mathrm{HCB}\cdot E_\mathrm{bonds}\\
   \phi_\mathrm{IM coating}\cdot E_\mathrm{tot}&=\phi_\mathrm{CIM}\cdot E_\mathrm{CIM}\\ \phi_\mathrm{FS}\cdot E_\mathrm{tot}&=\phi_\mathrm{Suprasil}\cdot E_\mathrm{FS}. 
   \end{split}
\end{align}
The overall loss angle for the IM is obtained by summing up all contributors:
$\phi_{\mathrm{IM}}=\phi_{\mathrm{Bonding}}+ \phi_{\mathrm{IM coating}}+\phi_{\mathrm{FS}}$. The mechanical quality factor of the IM modes then writes $Q_m=1/\phi_{\mathrm{IM}}$. 

Fig.~\ref{fig:phi_factors} shows the frequency dependence of the FS substrate loss and the effect of adding the optical coating and the bonding layers. The influence of the bonding term $\phi_{\mathrm{Bonding}}$ is strongly mode shape dependent through the deformation of the ear and anchor bulks and it is not negligible. In fact, its contribution to $Q_m$ is dominant. For this reason, from a set of Q measurements it is possible to infer the value of $\phi_{\mathrm{HCB}}$ by using the energy fractions calculated with the FEA.

Fig.~\ref{fig:q_factors} shows the $Q_m$ of the IM mass computed by fitting the loss angle $\phi_{\mathrm{HCB}}$ by using the first set of 5 modes of the IM of the north arm and supposing that it is does not vary with the frequency. At frequencies higher than 10~kHz, the bondings have a strong damping effect, though they have a negligible effect on the thermal noise of the IM. This is a very important result for the parametric gains computation and consequently for identification of the unstable modes.

\section{\label{sec:oms}Transverse optical modes in arm cavities}
Hermite-Gauss modes (HGM) are solutions of the paraxial wave equation for infinite-sized spherical mirrors. This mode basis was used in~\cite{evans2010general} to compute the parametric gain for the LIGO interferometer. It is fast to implement as the mode shapes are provided by analytical formulas. However, it restricts the mirror model to a purely spherical shape of infinite size. In particular, it does not include the effects of the deviations from the spherical shape due to fabrication imperfections or thermal effects. Finally, it does not take account for finite size effects such as diffraction losses, which must be estimated separately.

We have computed another set of optical modes that are obtained from a numerical resolution of the paraxial equation with finite-sized mirrors~\cite{kogelnik1966laser}. This mode basis will be referred to as `finite-sized mirror modes (FSMM)'. Contrary to HGM, FSMM are obtained directly with diffraction losses. Moreover, mirror shapes can be chosen arbitrary, which enables one to introduce any deformation of the mirrors due to thermal effects or fabrication imperfections. Note, that in this work, we did not include fabrication imperfections, which effects will be the subject of future work.

In the following, we analyze the differences in Gouy phase (or frequency), diffraction loss, and mode shape, and  between the HGM and FSMM basis set. 

\subsection{\label{sec:gouy_phases}Gouy phases}
\begin{figure}
	\centering
	\includegraphics[scale=1]{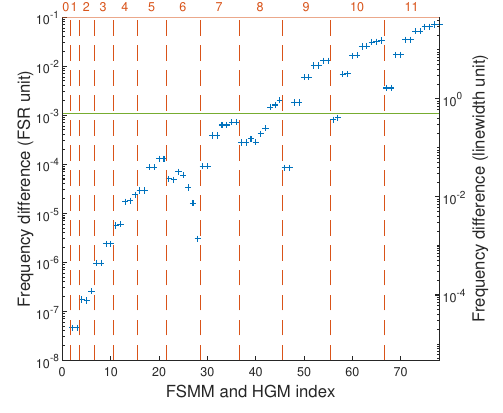}
	\caption{\label{fig:comparison_gouy_phases}Difference between HGM's and FSMM's Gouy phase, expressed in units of free spectral range on the left vertical axis, and in units of cavity linewidth on the right vertical axis. The vertical red dashed lines highlight the mode orders, which appear above the upper horizontal axis. The horizontal lower axis shows the optical mode index (modes are sorted by increasing energies). The green horizontal line has a vertical coordinate of 0.5 on the right axis.}
\end{figure}

The Gouy phases of arm cavity modes set the optical resonant frequencies, and, thus, the OPI resonance condition $\omega_m = \delta \omega$, where $\delta\omega$ is the difference in frequency between the zero order mode and the higher order transverse optical mode. In the case of HGM, the Gouy phase of the mode of linear index $n$ is given by
\begin{equation}
    \phi_{G_n} = O_n\,\phi_G \label{eq:gouy_phase}
\end{equation}
where $O_n$ the order of HGM(n), and $\phi_G$ is the Gouy phase of the lowest order mode HGM(1) (usually referred to as \TEMZZ~in the literature) given by
\begin{equation}
    \phi_G = \arccos{\left(-\sqrt{g_1\,g_2}\right)}.
\end{equation}
Here, $g_1<0$ and $g_2<0$ are the g parameters of the interferometer arm cavities. For the Advanced Virgo arm cavities, $\phi_G\simeq 2.74~$rad. and can be tuned by small variations of the mirror radii of curvature. Fig.~\ref{fig:comparison_gouy_phases} shows the difference between HGM's and FSMM's Gouy phases, expressed in units of free spectral range on the left vertical axis and in units of cavity linewidth on the vertical right axis. Note that the Gouy phases have been wrapped within an interval of length $\pi$, which allows to fold all the modes within a single free spectral range. The green line splits the graph into two regions: in the above region, the deviation is more than half a cavity linewidth, and we expect the model choice to have an impact on the OPI gain, whereas in the bottom part the impact should be negligible. Thus, the critical order is 7.

\subsection{\label{sec:diffraction_losses}Diffraction losses}
Diffraction losses stem from the finite size of the cavity mirrors. They are a key parameter to compute the parametric gain, since they contribute to the optical linewidth (together with material absorption losses, scattering losses, and mirror transmittance). Since low-order modes have most of the energy concentrated at the center of the mirror, their diffraction losses are small, whereas high-order modes spread over a larger surface and show higher diffraction losses. Thus, in general, high-order modes are less likely to contribute to a PI. However, note that counter-intuitively, a loss increase can sometimes lead to a higher parametric gain, as explained in more details in section~\ref{sec:cl_increase_rm}. 

\begin{figure}
	\centering
	\includegraphics[scale=1]{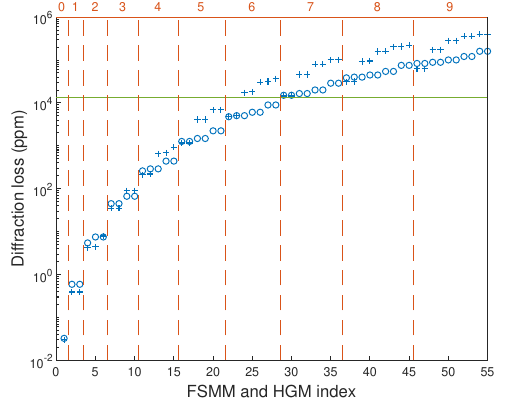}
	\caption{\label{fig:comparison_diffraction_losses}Diffraction losses obtained for FSMM (cross) and estimated for HGM (circles). The vertical red dashed lines underline the mode orders, which appear in the upper horizontal axis. The horizontal lower axis shows the optical mode index (modes are sorted by increasing energies). The green line shows the input mirror transmittance.}
\end{figure}

Diffraction losses for HGM are estimated, like in~\cite{evans2010general}, by evaluating the ratio between the total light flux within the coating radius of a mirror and the total flux incident on the mirror. Figure~\ref{fig:comparison_diffraction_losses} shows diffraction losses for both sets of modes. It shows that with this rough estimation method, besides the HGM(1), all HGM have their diffraction losses underestimated. However, note that the total losses (input mirror transmittance plus diffraction losses) of low-order modes are dominated by the input mirror transmittance (green line on Fig.~\ref{fig:comparison_diffraction_losses}). Thus, the total losses obtained with the two methods start to differ by more than 10\% around mode order 5.

\subsection{\label{sec:shapes}Mode amplitudes}
\begin{figure}
   \centering
   \includegraphics[scale=1]{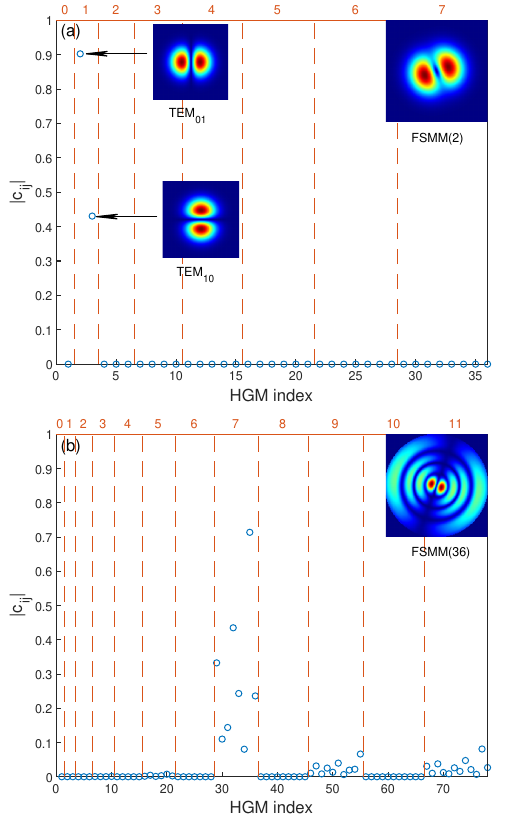}
\caption{\label{fig:comparison_decompositions}(a) $|c_{ij}|$ terms of the decomposition of the FSMM(2) on the HGM basis: the FSMM(2) is a linear combination of the HGM(2) and the HGM(3). (b) $|c_{ij}|$ terms of the decomposition of the FSMM(36) on the HGM basis.}
\end{figure}

Optical mode amplitudes are used to compute the three mode spatial overlap coefficient $B_{mn}$ of Eq.~\ref{OPIGain}. Thus, they also directly affect the OPI gain. In order to compare the mode amplitudes of FSMM and HGM, we decompose the vectors of one basis set onto the other by using the decomposition coefficient $c_{ij}$ of any FSMM (index $i$) with any HGM (index $j$):
\begin{equation}
c_{ij} =  \iint_{\mathcal{S}}\!\!\text{d}x\text{d}y ~ u_{i}^*(x, y)v_{\text{j}}(x, y),
\label{scalar_product}
\end{equation}
where $i$ and $j$ are modes integer indices, $u_i$ (resp. $v_j$) are the FSMM (resp. HGM) mode amplitudes, and $\mathcal{S}$ is the mirror coating surface. Note that the transverse profile of a FSMM is constrained on a disk (mirror coating), whereas for HGM the transverse profile is distributed over the whole plane, such that a linear superposition of FSMM will never exactly match a HGM, and a true transformation matrix between the two basis set cannot rigorously be obtained \cite{Siegman}. In Fig.~\ref{fig:comparison_decompositions}(a), we represent $|c_{ij}|$ for $i=2$, and $j\in \{1, 2, ..., 36\}$. We find that FSMM(2) is a linear combination of the two order one HGM which are the HGM(2) and the HGM(3). We find that this is true for orders below 7. Conversely, as shown in Fig.~\ref{fig:comparison_decompositions}(b), the higher order FSMM(36) mode (shown in the inset of Fig.~\ref{fig:comparison_decompositions}(b)) cannot be decomposed on a single order of HGM. In the presented case, it is a mixture of order 7, 9, 11, and many other higher odd orders that are not shown on the figure. 

\subsection{\label{sec:oms_conclusion}Conclusion}
This study shows that, in the absence of mirror deformation, the HGM basis does not deviate significantly from the FSMM basis for modes of order lower than 6. For order 6 and higher, the more resource consuming FSMM basis should lead to significantly different results for OPI gains. In section~\ref{sec:pi_computation}, we compare the OPI gains obtained for the Advanced Virgo O3 configuration, with HGM and FSMM basis set.

\begin{figure}[h!]
	\centering
	\includegraphics[scale=1]{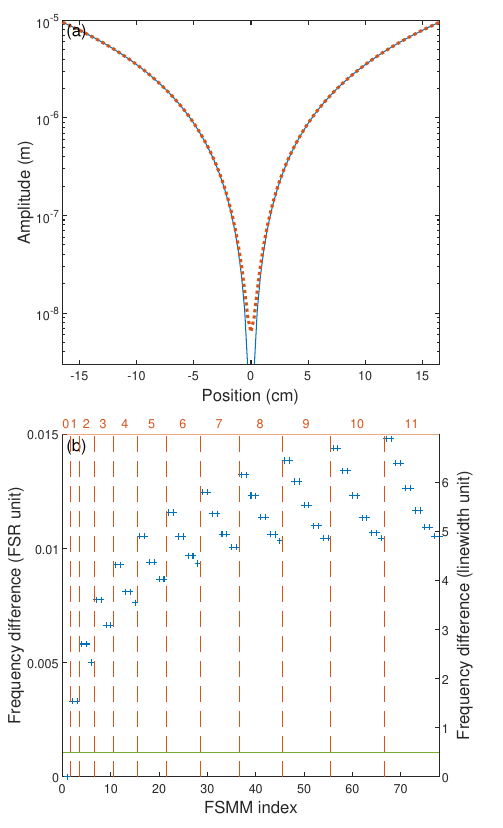}
	\caption{(a) Input mirror profiles with (dashed red line) and without thermal effect (blue solid line). (b) Frequency difference between FSMM modes with and without thermal effect, expressed in unit of FSR on the left axis, and in units of linewidth on the right axis. The green horizontal line is at vertical position 0.5 on the right axis.} 
	\label{fig:mir_and_thermal_deformation}
\end{figure}

\section{\label{sec:thermal_effects}Thermal effects}
The laser energy is partially absorbed both by coatings and in the bulk of mirrors. This causes a temperature gradient, which originates two effects. First, a gradient of refractive index in the bulk of input mirrors modifies the mode matching condition, but affects neither the cavity linewidth nor the mode frequencies. Second, a deformation of the mirror surface, which modifies the mode shapes and frequencies. In this part, we evaluate the impact on this second effect on the properties of cavity modes by comparing FSMM obtained for purely spherical mirrors with FSMM obtained for thermally deformed mirrors.

The deformation profile is obtained by solving the linear thermoelastic equations~\cite{virgo2005virgo}. Figure \ref{fig:mir_and_thermal_deformation}(a) shows the purely spherical and thermally deformed profiles of an Advanced Virgo input mirror, for an intracavity power of 300~kW. We fitted the central part of the deformed mirror to extract a radius of curvature. The results are given in the following table:

\begin{tabular}{|c|c|c|}
\hline
                 &  No thermal effect & With thermal effect\\ \hline
   Input mirror  & 1424.6 m            &   1432.1 m\\ \hline
   End mirror    & 1695 m              &   1702.3 m\\
\hline
\end{tabular}
\vspace{0.2cm}

However, note that the mirror is not spherical anymore and the result of the fit is only valid in the center. In order to evaluate the incidence of this effect on the optical cavity parameters, we computed the FSMM with and without thermal effect on the two cavity mirrors. Fig.~\ref{fig:mir_and_thermal_deformation}(b) shows the frequency differences between the two situations. We see that optical modes acquire a significantly different Gouy phase even for very low mode orders. We checked that losses and mode amplitudes are affected only for orders higher than 7, such that the frequency shift is the main effect. In section~\ref{sec:pi_computation}, we study the impact of this phenomenon on OPI gains.

\section{\label{sec:pi_computation}Parametric gain computation}

\subsection{\label{sec:validation} Validation: comparison with the Finesse software}

\begin{figure}[h!]
	\centering
	\includegraphics[scale=1]{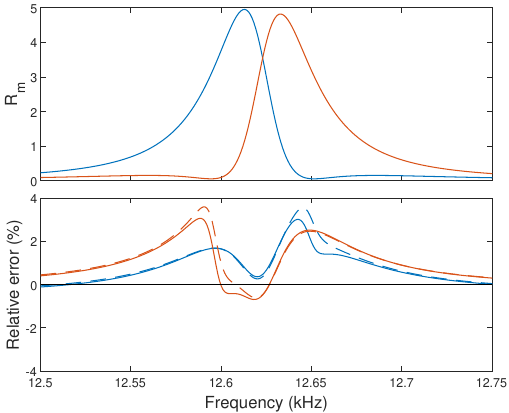}
\caption{\label{fig:rm_vs_cl}(a) Parametric gain obtained with the Finesse software for one mechanical mode (12.552~kHz) on mirror NE (blue) and WE (red). (b) Relative difference between (a) and the parametric gain $R_m$ obtained with Eq.~(\ref{OPIGain}) using FSMM (solid lines) and HGM (dashed lines).}
\label{fig_finesse}
\end{figure}

The OPI gains of all mechanical modes within the [5.7, 70.7]~kHz range were computed using Eq.~(\ref{OPIGain}) using both the HGM and FSMM basis set. In order to validate this method, we compared our results with the one obtained with the Finesse software~\cite{finesse_website,green2017influence}. The OPI gain obtained with the Finesse software for one mechanical mode and two arm cavity mirrors is shown in Fig.~\ref{fig_finesse}(a). In Fig.~\ref{fig_finesse}(b), we plot the relative difference with the OPI gain obtained with the Finesse software and with Eq.~(\ref{OPIGain}) using FSMM and HGM. We observe a difference of a few percent at maximum. Note that the slight asymmetry between the blue and red curves stems from the small parameter difference between the two arms cavities.  This comparison has been performed with many other mechanical modes and showed similar results. Note that using Eq.~(\ref{OPIGain}) is much faster than using the Finesse software, and that computing the results of the following figures would not have been possible in a reasonable amount of time. Therefore, in the following we use only Eq.~(\ref{OPIGain}).

\subsection{\label{sec:cl_increase_rm}Effect of optical losses on the OPI gain}
\begin{figure}
\centering
\includegraphics[scale=1]{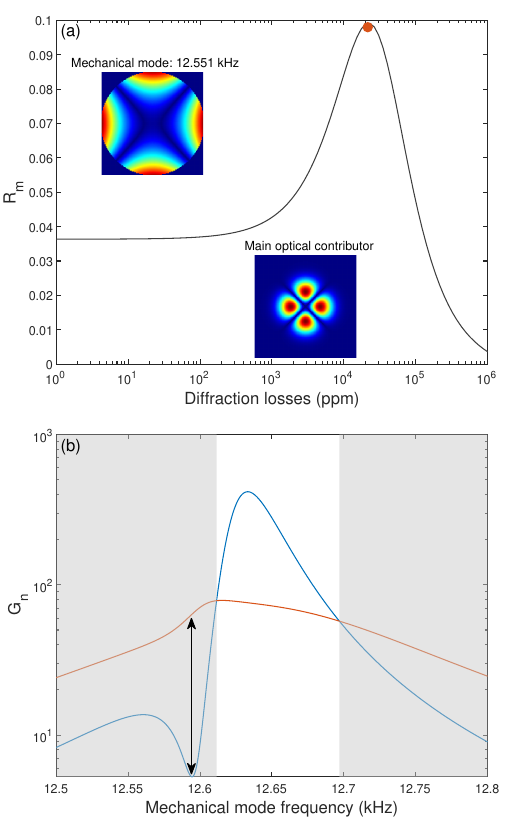}
\caption{(a) Parametric gain $R_m$ of a mechanical mode of frequency 12.5551~kHz while varying artificially the diffraction losses of optical modes. (b) Optical gain $G_n$ of a FSMM high order mode versus the mechanical mode frequency, which is artificially varied around the resonance condition $\delta\omega = \omega_m$ (with $\delta\omega =$ 12.5551~kHz). The blue line is for null diffraction losses, and the red one is for 21540~ppm (red bullet at the maximum of the curve in (a)). The arrow point the minimum in between the two resonance on the blue curve, where the gain increase is maximum. The gray shaded area highlights a frequency at which $G_n$ is higher when the diffraction loss is higher.}
\label{fig:impact_cl}
\end{figure}

In this section we demonstrate a counter-intuitive effect of optical losses on the OPI gains. Intuitively, if optical losses increase, the OPI gains get lower since the optical linewidth also increases. Here we show that if the OPI resonance condition is not exactly fulfilled, broadening the optical mode response can increase the gain such that counter intuitively, the gain variation does not vary monotonously with the diffraction loss. This is best shown on Fig.\ref{fig:impact_cl}(a), where the OPI gain of a mechanical mode is plotted against optical diffraction losses of the main optical contributor. In this example, the gain first increases from around 0.04 below $10^2$~ppm to 0.1 at $2\times 10^4$~ppm, before decreasing at higher loss values, as expected. This appears also in Fig.~\ref{fig:impact_cl}(b), where the optical gain $G_n$ of the main optical contributor to the OPI gain of the mechanical mode of Fig.~\ref{fig:impact_cl}(a) is represented as a function of the mechanical mode frequency, for two different values of diffraction losses. At low losses, the two resonance peaks are well separated, such that there is a minimum in between (black arrow). A loss increase from 0 to 21540~ppm (red bullet at maximum of Fig.~\ref{fig:impact_cl}(a)) broadens the peaks an lead to the red curve which has no minimum anymore, and which shows higher values in a whole frequency region (gray shaded area on Fig.~\ref{fig:impact_cl}(b)). Finally, if the losses were increased further, the red curve would start lowering and the gray shaded area would vanish.

\subsection{Impact of optical mode basis set on OPI gain}

\begin{figure}
	\centering
	\includegraphics[width=0.95\linewidth]{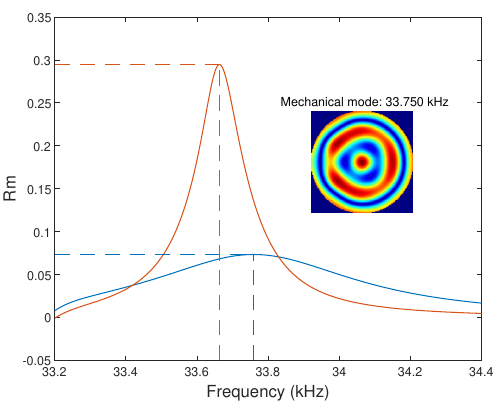}
	\caption{Parametric gain $R_m$ of the 33.750~kHz mechanical mode (see inset) while artificially varying the mechanical frequency, using the two different optical mode basis (red: HGM, blue: FSMM). The dashed lines point the maxima of the two curves, and emphasize the height and frequency change.}
	\label{fig:impact_om_basis}
\end{figure}

In this paragraph, we study the impact of the model used to compute the optical modes. We compare the OPI gains obtained with the HGM and FSMM basis. As expected, we find that there is only a marginal difference between the two models if optical modes of order below 5 are involved. In Fig.~\ref{fig:impact_om_basis}, we plot the gain of a mechanical mode versus its frequency using the two optical mode basis set. This mode has been chosen for the main optical contributor to the OPI gain is an order 6 optical mode. There is a factor 3 between the two gain maxima and the two peaks are shifted by around 100~Hz, which corresponds to the optical linewidth. This is in agreement with the conclusions of Sec.~\ref{sec:oms}.

\begin{table}
        \begin{tabular}{|l|c|l|}
        \hline
        Arm lengths                 & 2999.8~m\\
        Transmittance NI            & 13770~ppm\\
        Transmittance NE            & 4.4~ppm\\
        Transmittance WI            & 13750~ppm\\
        Transmittance WE            & 4.3~ppm\\
        Transmittance PR            & 48400~ppm\\
        Round trip loss             & 75~ppm\\
        Distance from BS to NI      & 6.0167~m\\
        Distance from BS to WI      & 5.7856~m\\
        Distance from BS to PR      & 6.0513~m\\
        Laser wavelength            & 1064~nm\\
        Gouy phase of PR cavity     & 1.8~mrad\\
       \hline
        \end{tabular}
 \caption{Advanced Virgo O3 optical parameters}
    \label{parametres}
\end{table}

\begin{figure*}
\includegraphics[scale=0.82]{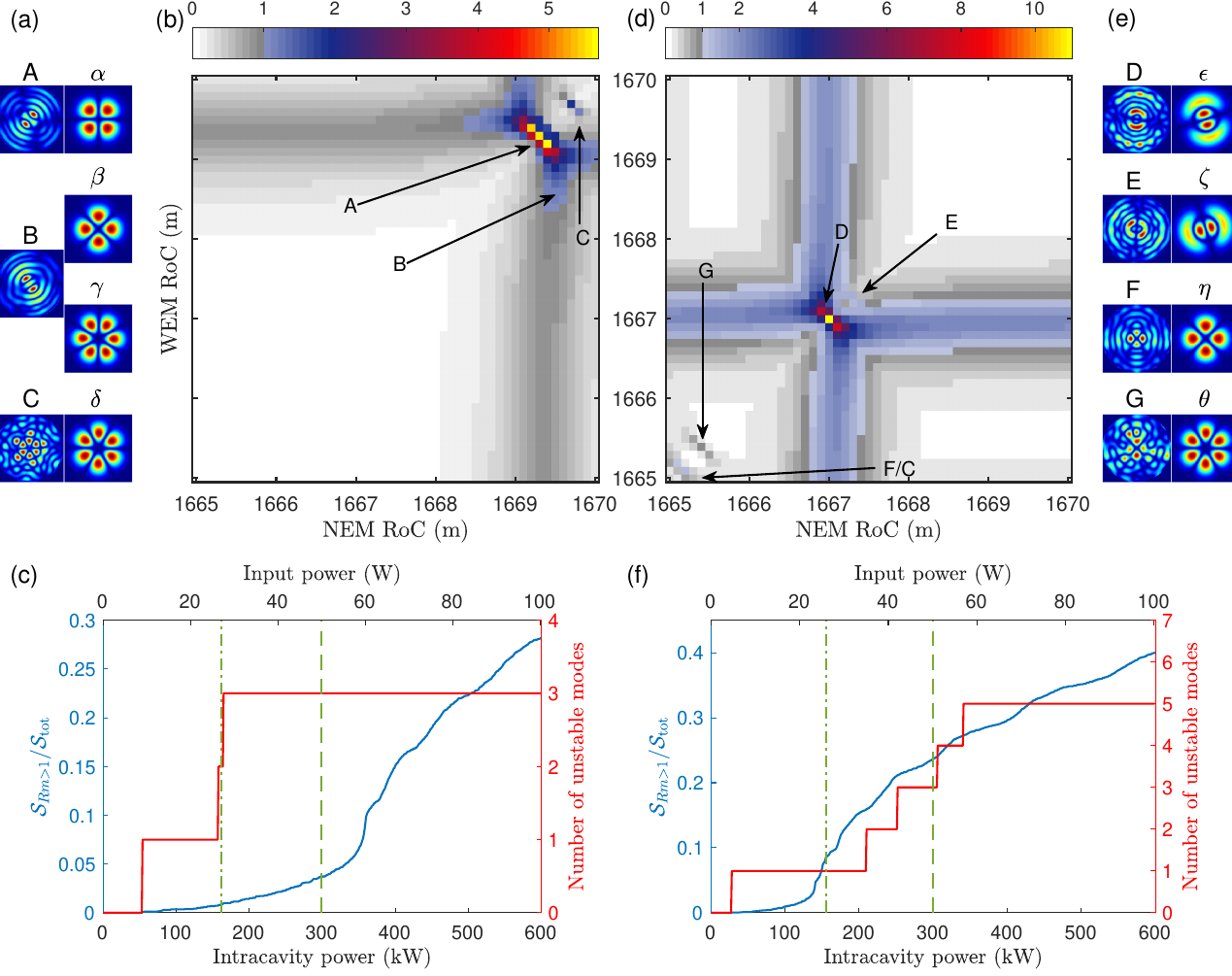}
\caption{(a) Parametric gain $R_m$ versus radii of curvature of NE and WE, using FSMM without thermal effect, for an input power of 50~W. The gray-shaded scale highlights the parametric gains lower than 1, while the colorful scale highlights the instabilities ($R_m > 1$). The three unstable mechanical modes are labeled A, B, and C. Their frequencies are respectively 61.154~kHz, 61.160~kHz, and 66.784~kHz. (b) The mechanical mode shapes and associated transverse optical mode contributing the most to the aforementioned instabilities. $\alpha$ contributes for 93~\% of A and for 77~\% of B, $\beta$ contributes for 6~\% of A and for 15~\% of B, $\gamma$ contributes for 89~\% of C, and $\delta$ for 10~\% of C. (c) The red solid line is the number of unstable modes in the radius of curvature range of (a), with respect to the optical input power (top) or intracavity power (bottom). They become unstable from an optical input power of respectively 9~W (mode A), 26~W (mode B), and 27~W (mode C). The blue curve is the ratio of the area free of instability $\mathcal{S}_{Rm > 1}$ to the total area $\mathcal{S}_{\rm{tot}}$ in Fig.~\ref{fig_gains} (a resp. b), versus the optical power. The figures (d) resp. (e) resp. (f) are the same than (a) resp. (b) resp. (c), but taking into account the thermal effect. The five unstable mechanical modes are labeled D, E, F, G, and C (same mode than without thermal effect). Their frequencies are respectively 66.888~kHz, 66.912~kHz, 61.216~kHz, 61.231~kHz, and 66.784~kHz. $\epsilon$ contributes for 84~\% of D and 24~\% of E, $\zeta$ contributes for 16~\% of D and 76~\% of E, $\eta$ contributes for 100~\% of F and G, and $\theta$ contributes for 97~\% of C. They become unstable from an optical input power of respectively 4.6~W (mode D), 35~W (mode E), 42~W (mode F), 51~W (mode G), and 57~W (mode C). The green dashed lines show the input power reached in O3 (28~W), and the nominal power (50~W) corresponding to (a) and (b).
}
\label{fig_gains}
\end{figure*}

\subsection{\label{sec:o3_prediction}OPI gain computation in the O3 configuration}

The simulations have been performed for the Advanced Virgo configuration corresponding to that of O3. The parameters for such a configuration are shown in Table~\ref{parametres}.  Measured parameters have been included rather than nominal values when they were available. Only the optical input power has been set to the nominal value of 50~W, which is the maximum value that would have been possibly used during O3 (the value effectively reached in O3 being too small to trigger any instability in the range of mechanical mode frequencies simulated in this work). The corresponding arm-cavity power is around 300~kW. The parameters used are listed in table~\ref{parametres}.

To account for optical mode frequency uncertainties, we present OPI data in two-dimensional plots (see for instance Fig.~\ref{fig_gains}(a) and (b)), where the end mirrors radii of curvature NE and WE are scanned. The color code indicates the gain value at each interferometer working point. This choice is also related to the fact that the envisioned OPI mitigation technique relies on ring heaters able to tune the end mirrors radii of curvatures~\cite{degallaix2007thermal}. In the following, we show two sets of results. Each figure is the result of the same OPI calculation but using a different set of optical modes. Fig.~\ref{fig_gains}(a, b, c) shows the results for FSMM, Fig.~\ref{fig_gains}(d, e, f) for FSMM including thermal effect due to coating absorption (see sec.~\ref{sec:thermal_effects}). NE and WE are scanned over a five meters range, which is within reach of the mirror ring heater system. In each OPI plot, the color code is chosen such that the gray scale is for gains lower than 1 (no instability), and the color scale is for $R_m>1$. The involved mechanical modes are indicated in the inset, and main optical modes contributors are shown below each OPI plot. Note that the result obtained with HGM are indistinguishable from that of Fig.~\ref{fig_gains}(a), such that we did not include the corresponding figure. Indeed, only low order optical modes are involved here, such that HGM and FSMM give the same result. Fig.~\ref{fig_gains}(b resp. d) shows the unstable mechanical modes of Fig.~\ref{fig_gains}(a resp. b) on the first line. The second line shows the involved optical modes. In Fig.~\ref{fig_gains}(c resp. f), we plot the number of unstable modes in the range of Fig.~\ref{fig_gains}(a resp. b)  versus the optical power for FSMM without (resp. with) thermal effect. Modes that ring on different mirrors are counted only once. The blue curves represents the ratio of the area free of instability $\mathcal{S}_{Rm > 1}$ to the total area $\mathcal{S}_{\rm{tot}}$ in Fig.~\ref{fig_gains} (a resp. b), versus the optical power. This ratio quantifies how difficult it is to escape an unstable area within the accessible radii of curvature range. The green vertical lines on Fig.~\ref{fig_gains}(c and f) point the nominal power of O3 (50~W) and the power that was effectively reached (28~W).

These results show that an OPI involving mechanical modes with frequencies below 70~kHz could have been observed at the nominal power of O3, although it would be easily escaped with end-mirrors ring heaters since $\mathcal{S}_{Rm > 1}/\mathcal{S}_{\rm{tot}}\simeq 0.2$ at 300~kW cavity power. It also shows that, at the real power of O3, it was very unlikely to observe an instability. Finally, it shows that the thermal effect caused by coating absorption has an important impact on the results at such high optical powers.

\section{\label{sec:conclusion}Conclusion}
In this letter, we have presented OPI gain simulations in the Advanced Virgo configuration of O3. Compared to previous work \cite{evans2010general}, we have used deeper physical modeling, including a very detailed description of mechanical modes and optical modes. The aim being that our model becomes predictive. The mechanical mode simulation includes all the mirrors details, and we implemented an original method to obtain precise quality factors value for all modes, by using the combination of FEA and ringdown measurement that are performed on a subset of modes. We have also provided a precise description of optical modes, by considering finite size effects. Our method provides directly accurate diffraction losses. Furthermore, we have shown that, counterintuitively, an optical loss increase can lead to a parametric gain increase.  Our conclusion regarding optical modes is that up to order 4 (included), analytical formulas for Hermite-Gauss modes are sufficient to predict accurate OPI gains. However, if higher-order optical modes are involved, mirrors finite size effects must be accounted for.  Finally, we have shown that the mirror deformation stemming from the laser absorption in mirror coatings plays an important role and must be included in the OPI simulation. These simulations pave the way towards precise optomechanical instability predictions for the current and next generations of gravitational-wave detectors.

\begin{acknowledgments}
We acknowledge Jérôme Degallaix, Ettore Majorana, François Bondu and Nelson Christensen for helping us to improve the manuscript. We thank Mikhaël Pichot Du Mezeray for providing us with the code to compute the mirror deformation
due to thermal effect, and Mourad Merzougui for discussions about Ansys calculations. We also thank Slawomir Gras, Jean-Yves Vinet and Walid Chaibi for interesting discussions and advice, Andreas Freise and Anna Green for their help with the Finesse software. Finally, David Cohen is supported by the DIM ACAV\texttt{+} from the Region Île-de-France.
\end{acknowledgments}

\bibliography{references}

\begin{thebibliography}{27}%
\makeatletter
\providecommand \@ifxundefined [1]{%
 \@ifx{#1\undefined}
}%
\providecommand \@ifnum [1]{%
 \ifnum #1\expandafter \@firstoftwo
 \else \expandafter \@secondoftwo
 \fi
}%
\providecommand \@ifx [1]{%
 \ifx #1\expandafter \@firstoftwo
 \else \expandafter \@secondoftwo
 \fi
}%
\providecommand \natexlab [1]{#1}%
\providecommand \enquote  [1]{``#1''}%
\providecommand \bibnamefont  [1]{#1}%
\providecommand \bibfnamefont [1]{#1}%
\providecommand \citenamefont [1]{#1}%
\providecommand \href@noop [0]{\@secondoftwo}%
\providecommand \href [0]{\begingroup \@sanitize@url \@href}%
\providecommand \@href[1]{\@@startlink{#1}\@@href}%
\providecommand \@@href[1]{\endgroup#1\@@endlink}%
\providecommand \@sanitize@url [0]{\catcode `\\12\catcode `\$12\catcode
  `\&12\catcode `\#12\catcode `\^12\catcode `\_12\catcode `\%12\relax}%
\providecommand \@@startlink[1]{}%
\providecommand \@@endlink[0]{}%
\providecommand \url  [0]{\begingroup\@sanitize@url \@url }%
\providecommand \@url [1]{\endgroup\@href {#1}{\urlprefix }}%
\providecommand \urlprefix  [0]{URL }%
\providecommand \Eprint [0]{\href }%
\providecommand \doibase [0]{https://doi.org/}%
\providecommand \selectlanguage [0]{\@gobble}%
\providecommand \bibinfo  [0]{\@secondoftwo}%
\providecommand \bibfield  [0]{\@secondoftwo}%
\providecommand \translation [1]{[#1]}%
\providecommand \BibitemOpen [0]{}%
\providecommand \bibitemStop [0]{}%
\providecommand \bibitemNoStop [0]{.\EOS\space}%
\providecommand \EOS [0]{\spacefactor3000\relax}%
\providecommand \BibitemShut  [1]{\csname bibitem#1\endcsname}%
\let\auto@bib@innerbib\@empty
\bibitem [{vir()}]{virgo_website}%
  \BibitemOpen
  \href {http://www.virgo-gw.eu} {\bibinfo {title}
  {http://www.virgo-gw.eu}}\BibitemShut {NoStop}%
\bibitem [{\citenamefont {Acernese}\ \emph {et~al.}(2014)\citenamefont
  {Acernese} \emph {et~al.}}]{acernese2014advanced}%
  \BibitemOpen
  \bibfield  {author} {\bibinfo {author} {\bibfnamefont {F.}~\bibnamefont
  {Acernese}} \emph {et~al.} (\bibinfo {collaboration} {The Virgo
  Collaboration}),\ }\href {https://doi.org/10.1088/0264-9381/32/2/024001}
  {\bibfield  {journal} {\bibinfo  {journal} {Classical and Quantum Gravity}\
  }\textbf {\bibinfo {volume} {32}},\ \bibinfo {pages} {024001} (\bibinfo
  {year} {2014})}\BibitemShut {NoStop}%
\bibitem [{lig()}]{ligo_website}%
  \BibitemOpen
  \href {https://www.ligo.caltech.edu} {\bibinfo {title}
  {https://www.ligo.caltech.edu}}\BibitemShut {NoStop}%
\bibitem [{\citenamefont {Aasi}\ \emph {et~al.}(2015)\citenamefont {Aasi} \emph
  {et~al.}}]{aasi2015advanced}%
  \BibitemOpen
  \bibfield  {author} {\bibinfo {author} {\bibfnamefont {J.}~\bibnamefont
  {Aasi}} \emph {et~al.} (\bibinfo {collaboration} {The LIGO Scientific
  Collaboration}),\ }\href {https://doi.org/10.1088/0264-9381/32/7/074001}
  {\bibfield  {journal} {\bibinfo  {journal} {Class. Quantum Grav.}\ }\textbf
  {\bibinfo {volume} {32}},\ \bibinfo {pages} {074001} (\bibinfo {year}
  {2015})}\BibitemShut {NoStop}%
\bibitem [{\citenamefont {Abbott}\ \emph
  {et~al.}(2016{\natexlab{a}})\citenamefont {Abbott} \emph
  {et~al.}}]{abbott2016observation}%
  \BibitemOpen
  \bibfield  {author} {\bibinfo {author} {\bibfnamefont {B.~P.}\ \bibnamefont
  {Abbott}} \emph {et~al.} (\bibinfo {collaboration} {LIGO Scientific
  Collaboration, Virgo Collaboration}),\ }\href
  {https://doi.org/10.1103/PhysRevLett.116.061102} {\bibfield  {journal}
  {\bibinfo  {journal} {Phys. Rev. Lett.}\ }\textbf {\bibinfo {volume} {116}},\
  \bibinfo {pages} {061102} (\bibinfo {year} {2016}{\natexlab{a}})}\BibitemShut
  {NoStop}%
\bibitem [{\citenamefont {Abbott}\ \emph
  {et~al.}(2019{\natexlab{a}})\citenamefont {Abbott} \emph
  {et~al.}}]{abbott2019gwtc}%
  \BibitemOpen
  \bibfield  {author} {\bibinfo {author} {\bibfnamefont {B.~P.}\ \bibnamefont
  {Abbott}} \emph {et~al.} (\bibinfo {collaboration} {LIGO Scientific
  Collaboration, Virgo Collaboration}),\ }\href
  {https://doi.org/10.1103/PhysRevX.9.031040} {\bibfield  {journal} {\bibinfo
  {journal} {Phys. Rev. X}\ }\textbf {\bibinfo {volume} {9}},\ \bibinfo {pages}
  {031040} (\bibinfo {year} {2019}{\natexlab{a}})}\BibitemShut {NoStop}%
\bibitem [{\citenamefont {Abbott}\ \emph {et~al.}(2020)\citenamefont {Abbott}
  \emph {et~al.}}]{abbott2020gwtc}%
  \BibitemOpen
  \bibfield  {author} {\bibinfo {author} {\bibfnamefont {R.}~\bibnamefont
  {Abbott}} \emph {et~al.} (\bibinfo {collaboration} {LIGO Scientific
  Collaboration, Virgo Collaboration}),\ }\href@noop {} {\bibfield  {journal}
  {\bibinfo  {journal} {arXiv preprint arXiv:2010.14527}\ } (\bibinfo {year}
  {2020})}\BibitemShut {NoStop}%
\bibitem [{\citenamefont {Abbott}\ \emph
  {et~al.}(2016{\natexlab{b}})\citenamefont {Abbott} \emph
  {et~al.}}]{abbott2016astrophysical}%
  \BibitemOpen
  \bibfield  {author} {\bibinfo {author} {\bibfnamefont {B.~P.}\ \bibnamefont
  {Abbott}} \emph {et~al.} (\bibinfo {collaboration} {LIGO Scientific
  Collaboration, Virgo Collaboration}),\ }\href
  {https://doi.org/10.3847/2041-8205/818/2/l22} {\bibfield  {journal} {\bibinfo
   {journal} {The Astrophysical Journal}\ }\textbf {\bibinfo {volume} {818}},\
  \bibinfo {pages} {L22} (\bibinfo {year} {2016}{\natexlab{b}})}\BibitemShut
  {NoStop}%
\bibitem [{\citenamefont {Abbott}\ \emph
  {et~al.}(2019{\natexlab{b}})\citenamefont {Abbott} \emph
  {et~al.}}]{abbott2019binary}%
  \BibitemOpen
  \bibfield  {author} {\bibinfo {author} {\bibfnamefont {B.~P.}\ \bibnamefont
  {Abbott}} \emph {et~al.} (\bibinfo {collaboration} {(LIGO Scientific
  Collaboration, Virgo Collaboration)}),\ }\href
  {https://doi.org/10.3847/2041-8213/ab3800} {\bibfield  {journal} {\bibinfo
  {journal} {The Astrophysical Journal}\ }\textbf {\bibinfo {volume} {882}},\
  \bibinfo {pages} {L24} (\bibinfo {year} {2019}{\natexlab{b}})}\BibitemShut
  {NoStop}%
\bibitem [{\citenamefont {Abbott}\ \emph
  {et~al.}(2017{\natexlab{a}})\citenamefont {Abbott} \emph
  {et~al.}}]{abbott2017gw170104}%
  \BibitemOpen
  \bibfield  {author} {\bibinfo {author} {\bibfnamefont {B.~P.}\ \bibnamefont
  {Abbott}} \emph {et~al.} (\bibinfo {collaboration} {LIGO Scientific
  Collaboration, Virgo Collaboration}),\ }\href
  {https://doi.org/10.1103/PhysRevLett.118.221101} {\bibfield  {journal}
  {\bibinfo  {journal} {Phys. Rev. Lett.}\ }\textbf {\bibinfo {volume} {118}},\
  \bibinfo {pages} {221101} (\bibinfo {year} {2017}{\natexlab{a}})}\BibitemShut
  {NoStop}%
\bibitem [{\citenamefont {Abbott}\ \emph
  {et~al.}(2017{\natexlab{b}})\citenamefont {Abbott} \emph
  {et~al.}}]{abbott2017gw170817}%
  \BibitemOpen
  \bibfield  {author} {\bibinfo {author} {\bibfnamefont {B.~P.}\ \bibnamefont
  {Abbott}} \emph {et~al.} (\bibinfo {collaboration} {LIGO Scientific
  Collaboration, Virgo Collaboration}),\ }\href
  {https://doi.org/10.1103/PhysRevLett.119.161101} {\bibfield  {journal}
  {\bibinfo  {journal} {Phys. Rev. Lett.}\ }\textbf {\bibinfo {volume} {119}},\
  \bibinfo {pages} {161101} (\bibinfo {year} {2017}{\natexlab{b}})}\BibitemShut
  {NoStop}%
\bibitem [{\citenamefont {Abbott}\ \emph
  {et~al.}(2017{\natexlab{c}})\citenamefont {Abbott} \emph
  {et~al.}}]{ligo2017gravitational}%
  \BibitemOpen
  \bibfield  {author} {\bibinfo {author} {\bibfnamefont {B.~P.}\ \bibnamefont
  {Abbott}} \emph {et~al.} (\bibinfo {collaboration} {LIGO Scientific
  Collaboration, Virgo Collaboration, 1M2H Collaboration, Dark Energy Camera
  GW-EM Collaboration, DES Collaboration, DLT40 Collaboration, Las Cumbres
  Observatory Collaboration, VINROUGE Collaboration, MASTER Collaboration}),\
  }\href {https://doi.org/https://doi.org/10.1038/nature24471} {\bibfield
  {journal} {\bibinfo  {journal} {Nature (London)}\ }\textbf {\bibinfo {volume}
  {551}},\ \bibinfo {pages} {85} (\bibinfo {year}
  {2017}{\natexlab{c}})}\BibitemShut {NoStop}%
\bibitem [{\citenamefont {Braginsky}\ \emph {et~al.}(2001)\citenamefont
  {Braginsky}, \citenamefont {Strigin},\ and\ \citenamefont
  {Vyatchanin}}]{braginsky2001parametric}%
  \BibitemOpen
  \bibfield  {author} {\bibinfo {author} {\bibfnamefont {V.}~\bibnamefont
  {Braginsky}}, \bibinfo {author} {\bibfnamefont {S.}~\bibnamefont {Strigin}},\
  and\ \bibinfo {author} {\bibfnamefont {S.}~\bibnamefont {Vyatchanin}},\
  }\href {https://doi.org/https://doi.org/10.1016/S0375-9601(01)00510-2}
  {\bibfield  {journal} {\bibinfo  {journal} {Physics Letters A}\ }\textbf
  {\bibinfo {volume} {287}},\ \bibinfo {pages} {331} (\bibinfo {year}
  {2001})}\BibitemShut {NoStop}%
\bibitem [{\citenamefont {Vyatchanin}\ and\ \citenamefont
  {Strigin}(2012)}]{vyatchanin2012parametric}%
  \BibitemOpen
  \bibfield  {author} {\bibinfo {author} {\bibfnamefont {S.~P.}\ \bibnamefont
  {Vyatchanin}}\ and\ \bibinfo {author} {\bibfnamefont {S.~E.}\ \bibnamefont
  {Strigin}},\ }\href {https://doi.org/10.3367/ufne.0182.201211e.1195}
  {\bibfield  {journal} {\bibinfo  {journal} {Physics-Uspekhi}\ }\textbf
  {\bibinfo {volume} {55}},\ \bibinfo {pages} {1115} (\bibinfo {year}
  {2012})}\BibitemShut {NoStop}%
\bibitem [{\citenamefont {Evans}\ \emph {et~al.}(2015)\citenamefont {Evans},
  \citenamefont {Gras}, \citenamefont {Fritschel}, \citenamefont {Miller},
  \citenamefont {Barsotti}, \citenamefont {Martynov}, \citenamefont {Brooks},
  \citenamefont {Coyne}, \citenamefont {Abbott}, \citenamefont {Adhikari},
  \citenamefont {Arai}, \citenamefont {Bork}, \citenamefont {Kells},
  \citenamefont {Rollins}, \citenamefont {Smith-Lefebvre}, \citenamefont
  {Vajente}, \citenamefont {Yamamoto}, \citenamefont {Adams}, \citenamefont
  {Aston}, \citenamefont {Betzweiser}, \citenamefont {Frolov}, \citenamefont
  {Mullavey}, \citenamefont {Pele}, \citenamefont {Romie}, \citenamefont
  {Thomas}, \citenamefont {Thorne}, \citenamefont {Dwyer}, \citenamefont
  {Izumi}, \citenamefont {Kawabe}, \citenamefont {Sigg}, \citenamefont
  {Derosa}, \citenamefont {Effler}, \citenamefont {Kokeyama}, \citenamefont
  {Ballmer}, \citenamefont {Massinger}, \citenamefont {Staley}, \citenamefont
  {Heinze}, \citenamefont {Mueller}, \citenamefont {Grote}, \citenamefont
  {Ward}, \citenamefont {King}, \citenamefont {Blair}, \citenamefont {Ju},\
  and\ \citenamefont {Zhao}}]{evans2015observation}%
  \BibitemOpen
  \bibfield  {author} {\bibinfo {author} {\bibfnamefont {M.}~\bibnamefont
  {Evans}}, \bibinfo {author} {\bibfnamefont {S.}~\bibnamefont {Gras}},
  \bibinfo {author} {\bibfnamefont {P.}~\bibnamefont {Fritschel}}, \bibinfo
  {author} {\bibfnamefont {J.}~\bibnamefont {Miller}}, \bibinfo {author}
  {\bibfnamefont {L.}~\bibnamefont {Barsotti}}, \bibinfo {author}
  {\bibfnamefont {D.}~\bibnamefont {Martynov}}, \bibinfo {author}
  {\bibfnamefont {A.}~\bibnamefont {Brooks}}, \bibinfo {author} {\bibfnamefont
  {D.}~\bibnamefont {Coyne}}, \bibinfo {author} {\bibfnamefont
  {R.}~\bibnamefont {Abbott}}, \bibinfo {author} {\bibfnamefont {R.~X.}\
  \bibnamefont {Adhikari}}, \bibinfo {author} {\bibfnamefont {K.}~\bibnamefont
  {Arai}}, \bibinfo {author} {\bibfnamefont {R.}~\bibnamefont {Bork}}, \bibinfo
  {author} {\bibfnamefont {B.}~\bibnamefont {Kells}}, \bibinfo {author}
  {\bibfnamefont {J.}~\bibnamefont {Rollins}}, \bibinfo {author} {\bibfnamefont
  {N.}~\bibnamefont {Smith-Lefebvre}}, \bibinfo {author} {\bibfnamefont
  {G.}~\bibnamefont {Vajente}}, \bibinfo {author} {\bibfnamefont
  {H.}~\bibnamefont {Yamamoto}}, \bibinfo {author} {\bibfnamefont
  {C.}~\bibnamefont {Adams}}, \bibinfo {author} {\bibfnamefont
  {S.}~\bibnamefont {Aston}}, \bibinfo {author} {\bibfnamefont
  {J.}~\bibnamefont {Betzweiser}}, \bibinfo {author} {\bibfnamefont
  {V.}~\bibnamefont {Frolov}}, \bibinfo {author} {\bibfnamefont
  {A.}~\bibnamefont {Mullavey}}, \bibinfo {author} {\bibfnamefont
  {A.}~\bibnamefont {Pele}}, \bibinfo {author} {\bibfnamefont {J.}~\bibnamefont
  {Romie}}, \bibinfo {author} {\bibfnamefont {M.}~\bibnamefont {Thomas}},
  \bibinfo {author} {\bibfnamefont {K.}~\bibnamefont {Thorne}}, \bibinfo
  {author} {\bibfnamefont {S.}~\bibnamefont {Dwyer}}, \bibinfo {author}
  {\bibfnamefont {K.}~\bibnamefont {Izumi}}, \bibinfo {author} {\bibfnamefont
  {K.}~\bibnamefont {Kawabe}}, \bibinfo {author} {\bibfnamefont
  {D.}~\bibnamefont {Sigg}}, \bibinfo {author} {\bibfnamefont {R.}~\bibnamefont
  {Derosa}}, \bibinfo {author} {\bibfnamefont {A.}~\bibnamefont {Effler}},
  \bibinfo {author} {\bibfnamefont {K.}~\bibnamefont {Kokeyama}}, \bibinfo
  {author} {\bibfnamefont {S.}~\bibnamefont {Ballmer}}, \bibinfo {author}
  {\bibfnamefont {T.~J.}\ \bibnamefont {Massinger}}, \bibinfo {author}
  {\bibfnamefont {A.}~\bibnamefont {Staley}}, \bibinfo {author} {\bibfnamefont
  {M.}~\bibnamefont {Heinze}}, \bibinfo {author} {\bibfnamefont
  {C.}~\bibnamefont {Mueller}}, \bibinfo {author} {\bibfnamefont
  {H.}~\bibnamefont {Grote}}, \bibinfo {author} {\bibfnamefont
  {R.}~\bibnamefont {Ward}}, \bibinfo {author} {\bibfnamefont {E.}~\bibnamefont
  {King}}, \bibinfo {author} {\bibfnamefont {D.}~\bibnamefont {Blair}},
  \bibinfo {author} {\bibfnamefont {L.}~\bibnamefont {Ju}},\ and\ \bibinfo
  {author} {\bibfnamefont {C.}~\bibnamefont {Zhao}},\ }\href
  {https://doi.org/10.1103/PhysRevLett.114.161102} {\bibfield  {journal}
  {\bibinfo  {journal} {Phys. Rev. Lett.}\ }\textbf {\bibinfo {volume} {114}},\
  \bibinfo {pages} {161102} (\bibinfo {year} {2015})}\BibitemShut {NoStop}%
\bibitem [{\citenamefont {Chen}\ \emph {et~al.}(2015)\citenamefont {Chen},
  \citenamefont {Zhao}, \citenamefont {Danilishin}, \citenamefont {Ju},
  \citenamefont {Blair}, \citenamefont {Wang}, \citenamefont {Vyatchanin},
  \citenamefont {Molinelli}, \citenamefont {Kuhn}, \citenamefont {Gras},
  \citenamefont {Briant}, \citenamefont {Cohadon}, \citenamefont {Heidmann},
  \citenamefont {Roch-Jeune}, \citenamefont {Flaminio}, \citenamefont
  {Michel},\ and\ \citenamefont {Pinard}}]{Chen}%
  \BibitemOpen
  \bibfield  {author} {\bibinfo {author} {\bibfnamefont {X.}~\bibnamefont
  {Chen}}, \bibinfo {author} {\bibfnamefont {C.}~\bibnamefont {Zhao}}, \bibinfo
  {author} {\bibfnamefont {S.}~\bibnamefont {Danilishin}}, \bibinfo {author}
  {\bibfnamefont {L.}~\bibnamefont {Ju}}, \bibinfo {author} {\bibfnamefont
  {D.}~\bibnamefont {Blair}}, \bibinfo {author} {\bibfnamefont
  {H.}~\bibnamefont {Wang}}, \bibinfo {author} {\bibfnamefont {S.~P.}\
  \bibnamefont {Vyatchanin}}, \bibinfo {author} {\bibfnamefont
  {C.}~\bibnamefont {Molinelli}}, \bibinfo {author} {\bibfnamefont
  {A.}~\bibnamefont {Kuhn}}, \bibinfo {author} {\bibfnamefont {S.}~\bibnamefont
  {Gras}}, \bibinfo {author} {\bibfnamefont {T.}~\bibnamefont {Briant}},
  \bibinfo {author} {\bibfnamefont {P.-F.}\ \bibnamefont {Cohadon}}, \bibinfo
  {author} {\bibfnamefont {A.}~\bibnamefont {Heidmann}}, \bibinfo {author}
  {\bibfnamefont {I.}~\bibnamefont {Roch-Jeune}}, \bibinfo {author}
  {\bibfnamefont {R.}~\bibnamefont {Flaminio}}, \bibinfo {author}
  {\bibfnamefont {C.}~\bibnamefont {Michel}},\ and\ \bibinfo {author}
  {\bibfnamefont {L.}~\bibnamefont {Pinard}},\ }\href
  {https://doi.org/10.1103/PhysRevA.91.033832} {\bibfield  {journal} {\bibinfo
  {journal} {Phys. Rev. A}\ }\textbf {\bibinfo {volume} {91}},\ \bibinfo
  {pages} {033832} (\bibinfo {year} {2015})}\BibitemShut {NoStop}%
\bibitem [{\citenamefont {Evans}\ \emph {et~al.}(2010)\citenamefont {Evans},
  \citenamefont {Barsotti},\ and\ \citenamefont
  {Fritschel}}]{evans2010general}%
  \BibitemOpen
  \bibfield  {author} {\bibinfo {author} {\bibfnamefont {M.}~\bibnamefont
  {Evans}}, \bibinfo {author} {\bibfnamefont {L.}~\bibnamefont {Barsotti}},\
  and\ \bibinfo {author} {\bibfnamefont {P.}~\bibnamefont {Fritschel}},\ }\href
  {https://doi.org/https://doi.org/10.1016/j.physleta.2009.11.023} {\bibfield
  {journal} {\bibinfo  {journal} {Physics Letters A}\ }\textbf {\bibinfo
  {volume} {374}},\ \bibinfo {pages} {665} (\bibinfo {year}
  {2010})}\BibitemShut {NoStop}%
\bibitem [{\citenamefont {Gras}\ \emph {et~al.}(2010)\citenamefont {Gras},
  \citenamefont {Zhao}, \citenamefont {Blair},\ and\ \citenamefont
  {Ju}}]{Grass}%
  \BibitemOpen
  \bibfield  {author} {\bibinfo {author} {\bibfnamefont {S.}~\bibnamefont
  {Gras}}, \bibinfo {author} {\bibfnamefont {C.}~\bibnamefont {Zhao}}, \bibinfo
  {author} {\bibfnamefont {D.~G.}\ \bibnamefont {Blair}},\ and\ \bibinfo
  {author} {\bibfnamefont {L.}~\bibnamefont {Ju}},\ }\href
  {https://doi.org/10.1088/0264-9381/27/20/205019} {\bibfield  {journal}
  {\bibinfo  {journal} {Classical and Quantum Gravity}\ }\textbf {\bibinfo
  {volume} {27}},\ \bibinfo {pages} {205019} (\bibinfo {year}
  {2010})}\BibitemShut {NoStop}%
\bibitem [{cun(2010)}]{cunningham}%
  \BibitemOpen
  \href {https://doi.org/https://doi.org/10.1016/j.physleta.2010.07.049}
  {\bibfield  {journal} {\bibinfo  {journal} {Physics Letters A}\ }\textbf
  {\bibinfo {volume} {374}},\ \bibinfo {pages} {3993} (\bibinfo {year}
  {2010})}\BibitemShut {NoStop}%
\bibitem [{\citenamefont {Granata}\ \emph {et~al.}(2020)\citenamefont
  {Granata}, \citenamefont {Amato}, \citenamefont {Balzarini}, \citenamefont
  {Canepa}, \citenamefont {Degallaix}, \citenamefont {Forest}, \citenamefont
  {Dolique}, \citenamefont {Mereni}, \citenamefont {Michel}, \citenamefont
  {Pinard}, \citenamefont {Sassolas}, \citenamefont {Teillon},\ and\
  \citenamefont {Cagnoli}}]{granata}%
  \BibitemOpen
  \bibfield  {author} {\bibinfo {author} {\bibfnamefont {M.}~\bibnamefont
  {Granata}}, \bibinfo {author} {\bibfnamefont {A.}~\bibnamefont {Amato}},
  \bibinfo {author} {\bibfnamefont {L.}~\bibnamefont {Balzarini}}, \bibinfo
  {author} {\bibfnamefont {M.}~\bibnamefont {Canepa}}, \bibinfo {author}
  {\bibfnamefont {J.}~\bibnamefont {Degallaix}}, \bibinfo {author}
  {\bibfnamefont {D.}~\bibnamefont {Forest}}, \bibinfo {author} {\bibfnamefont
  {V.}~\bibnamefont {Dolique}}, \bibinfo {author} {\bibfnamefont
  {L.}~\bibnamefont {Mereni}}, \bibinfo {author} {\bibfnamefont
  {C.}~\bibnamefont {Michel}}, \bibinfo {author} {\bibfnamefont
  {L.}~\bibnamefont {Pinard}}, \bibinfo {author} {\bibfnamefont
  {B.}~\bibnamefont {Sassolas}}, \bibinfo {author} {\bibfnamefont
  {J.}~\bibnamefont {Teillon}},\ and\ \bibinfo {author} {\bibfnamefont
  {G.}~\bibnamefont {Cagnoli}},\ }\href
  {https://doi.org/10.1088/1361-6382/ab77e9} {\bibfield  {journal} {\bibinfo
  {journal} {Classical and Quantum Gravity}\ }\textbf {\bibinfo {volume}
  {37}},\ \bibinfo {pages} {095004} (\bibinfo {year} {2020})}\BibitemShut
  {NoStop}%
\bibitem [{\citenamefont {Puppo}(2019)}]{puppo}%
  \BibitemOpen
  \bibfield  {author} {\bibinfo {author} {\bibfnamefont {P.}~\bibnamefont
  {Puppo}},\ }\href {https://tds.virgo-gw.eu/ql/?c=12425} {\bibinfo {title}
  {Evaluation of the quality factors of the mirrors in advanced virgo using the
  fem ansys model, https://tds.virgo-gw.eu/ql/?c=12425}} (\bibinfo {year}
  {2019})\BibitemShut {NoStop}%
\bibitem [{\citenamefont {Kogelnik}\ and\ \citenamefont
  {Li}(1966)}]{kogelnik1966laser}%
  \BibitemOpen
  \bibfield  {author} {\bibinfo {author} {\bibfnamefont {H.}~\bibnamefont
  {Kogelnik}}\ and\ \bibinfo {author} {\bibfnamefont {T.}~\bibnamefont {Li}},\
  }\href {https://doi.org/10.1364/AO.5.001550} {\bibfield  {journal} {\bibinfo
  {journal} {Appl. Opt.}\ }\textbf {\bibinfo {volume} {5}},\ \bibinfo {pages}
  {1550} (\bibinfo {year} {1966})}\BibitemShut {NoStop}%
\bibitem [{\citenamefont {Siegman}(1986)}]{Siegman}%
  \BibitemOpen
  \bibfield  {author} {\bibinfo {author} {\bibfnamefont {A.~E.}\ \bibnamefont
  {Siegman}},\ }\href {https://books.google.fr/books?id=qBBaDQEACAAJ} {\emph
  {\bibinfo {title} {Lasers}}}\ (\bibinfo  {publisher} {University Science
  Books},\ \bibinfo {year} {1986})\BibitemShut {NoStop}%
\bibitem [{\citenamefont {Vinet}(2020)}]{virgo2005virgo}%
  \BibitemOpen
  \bibfield  {author} {\bibinfo {author} {\bibfnamefont {J.-Y.}\ \bibnamefont
  {Vinet}},\ }\href {https://artemis.oca.eu/VinetOptics} {\bibinfo {title} {The
  virgo physics book: Optics and related topics, chap. 6}} (\bibinfo {year}
  {2020}),\ \bibinfo {note} {{The book can be found at
  https://artemis.oca.eu/VinetOptics}}\BibitemShut {NoStop}%
\bibitem [{\citenamefont {Brown}\ and\ \citenamefont
  {Freise}(2014)}]{finesse_website}%
  \BibitemOpen
  \bibfield  {author} {\bibinfo {author} {\bibfnamefont {D.~D.}\ \bibnamefont
  {Brown}}\ and\ \bibinfo {author} {\bibfnamefont {A.}~\bibnamefont {Freise}},\
  }\href {https://doi.org/10.5281/zenodo.821363} {\bibinfo {title} {Finesse}}
  (\bibinfo {year} {2014}),\ \bibinfo {note} {{The software and source code are
  available at \url{http://www.gwoptics.org/finesse}.}}\BibitemShut {Stop}%
\bibitem [{\citenamefont {Green}\ \emph {et~al.}(2017)\citenamefont {Green},
  \citenamefont {Brown}, \citenamefont {Dovale-{\'{A}}lvarez}, \citenamefont
  {Collins}, \citenamefont {Miao}, \citenamefont {Mow-Lowry},\ and\
  \citenamefont {Freise}}]{green2017influence}%
  \BibitemOpen
  \bibfield  {author} {\bibinfo {author} {\bibfnamefont {A.~C.}\ \bibnamefont
  {Green}}, \bibinfo {author} {\bibfnamefont {D.~D.}\ \bibnamefont {Brown}},
  \bibinfo {author} {\bibfnamefont {M.}~\bibnamefont {Dovale-{\'{A}}lvarez}},
  \bibinfo {author} {\bibfnamefont {C.}~\bibnamefont {Collins}}, \bibinfo
  {author} {\bibfnamefont {H.}~\bibnamefont {Miao}}, \bibinfo {author}
  {\bibfnamefont {C.~M.}\ \bibnamefont {Mow-Lowry}},\ and\ \bibinfo {author}
  {\bibfnamefont {A.}~\bibnamefont {Freise}},\ }\href
  {https://doi.org/10.1088/1361-6382/aa8af8} {\bibfield  {journal} {\bibinfo
  {journal} {Classical and Quantum Gravity}\ }\textbf {\bibinfo {volume}
  {34}},\ \bibinfo {pages} {205004} (\bibinfo {year} {2017})}\BibitemShut
  {NoStop}%
\bibitem [{\citenamefont {Degallaix}\ \emph {et~al.}(2007)\citenamefont
  {Degallaix}, \citenamefont {Zhao}, \citenamefont {Ju},\ and\ \citenamefont
  {Blair}}]{degallaix2007thermal}%
  \BibitemOpen
  \bibfield  {author} {\bibinfo {author} {\bibfnamefont {J.}~\bibnamefont
  {Degallaix}}, \bibinfo {author} {\bibfnamefont {C.}~\bibnamefont {Zhao}},
  \bibinfo {author} {\bibfnamefont {L.}~\bibnamefont {Ju}},\ and\ \bibinfo
  {author} {\bibfnamefont {D.}~\bibnamefont {Blair}},\ }\href
  {https://doi.org/10.1364/JOSAB.24.001336} {\bibfield  {journal} {\bibinfo
  {journal} {J. Opt. Soc. Am. B}\ }\textbf {\bibinfo {volume} {24}},\ \bibinfo
  {pages} {1336} (\bibinfo {year} {2007})}\BibitemShut {NoStop}%
\end{thebibliography}%

\end{document}